\documentclass[twocolumn,trackchanges]{aastex631}
\usepackage[utf8]{inputenc}
\hypersetup{linkcolor=red,citecolor=blue,filecolor=cyan,urlcolor=magenta}

\usepackage{color}
\usepackage{xspace}
\usepackage{multirow}
\usepackage{ulem}

\newcommand{\Msun}{{\ensuremath{\mathrm{M_{\odot}}}}\xspace}
\newcommand{\Rsun}{{\ensuremath{\mathrm{R_{\odot}}}}\xspace}
\newcommand{\Lsun}{{\ensuremath{\mathrm{L_{\odot}}}}\xspace}
\newcommand{\Mch}{{\ensuremath{\mathrm{M_{\mathrm{Ch}}}}}\xspace}
\newcommand{\K}{{\ensuremath{\mathrm{K}}}\xspace}
\newcommand{\yr}{{\ensuremath{\mathrm{yr}}}\xspace}
\newcommand{\mesa}{{\textsc{MESA}}\xspace}
\newcommand{\kepler}{{\textsc{Kepler}}\xspace}

\shorttitle{Surviving He companions of SNe Iax}
\shortauthors{Y.~Zeng et al.}

\begin{document}
\title{Long-term evolution of post-explosion Helium-star Companions of Type Iax Supernovae}

\correspondingauthor{Zheng-Wei Liu}
\email{zwliu@ynao.ac.cn}

\author[0000-0003-1356-8642]{Yaotian Zeng}
\affiliation{Yunnan Observatories, Chinese Academy of Sciences (CAS), Kunming 650216, P.R.~China}
\affiliation{Key Laboratory for the Structure and Evolution of Celestial Objects, CAS, Kunming 650216, P.R.~China}
\affiliation{University of Chinese Academy of Science, Beijing 100012, P.R.~China}

\author[0000-0002-7909-4171]{Zheng-Wei Liu}
\affiliation{Yunnan Observatories, Chinese Academy of Sciences (CAS), Kunming 650216, P.R.~China}
\affiliation{Key Laboratory for the Structure and Evolution of Celestial Objects, CAS, Kunming 650216, P.R.~China}
\affiliation{University of Chinese Academy of Science, Beijing 100012, P.R.~China}
\affiliation{Center for Astronomical Mega-Science, CAS, Beijing 100012, P.R.~China}

\author[0000-0002-3684-1325]{Alexander Heger}
\affiliation{School of Physics and Astronomy, Monash University, 19 Rainforest Walk, VIC 3800, Australia}
\affiliation{Joint Institute for Nuclear Astrophysics, National Superconducting Cyclotron Laboratory, Michigan State University, 1 Cyclotron Laboratory, East Lansing,
MI 48824-1321, USA}
\affiliation{Australian Research Council Centre of Excellence for Gravitational Wave Discovery (OzGrav), Clayton, VIC 3800, Australia}
\affiliation{Center of Excellence for Astrophysics in Three Dimensions (ASTRO-3D), Stromlo, ACT 2611, Australia}

\author[0000-0001-5807-7893]{Curtis McCully}
\affiliation{Las Cumbres Observatory, 6740 Cortona Dr Ste 102, Goleta, CA 93117-5575, USA}
\affiliation{Department of Physics, University of California, Santa Barbara, CA 93106-9530, USA}

\author[0000-0002-4460-0097]{Friedrich K. R{\"o}pke}
\affiliation{Zentrum f{\"u}r Astronomie der Universit{\"a}t Heidelberg, Institut f{\"u}r Theoretische Astrophysik, Philosophenweg 12, D-69120 Heidelberg, Germany}
\affiliation{Heidelberger Institut f{\"u}r Theoretische Studien, Schloss-Wolfsbrunnenweg 35, D-69118 Heidelberg, Germany}

\author[0000-0001-9204-7778]{Zhanwen Han}
\affiliation{Yunnan Observatories, Chinese Academy of Sciences (CAS), Kunming 650216, P.R.~China}
\affiliation{Key Laboratory for the Structure and Evolution of Celestial Objects, CAS, Kunming 650216, P.R.~China}
\affiliation{University of Chinese Academy of Science, Beijing 100012, P.R.~China}
\affiliation{Center for Astronomical Mega-Science, CAS, Beijing 100012, P.R.~China}

\collaboration{6}{}

\begin{abstract}
Supernovae of Type Iax (SNe Iax) are an accepted faint subclass of hydrogen-free supernovae.  Their origin, the nature of the progenitor systems, however, is an open question.  Recent studies suggest that the weak deflagration explosion of a near-Chandrasekhar-mass white dwarf in a binary system with a helium star donor could be the origin of SNe Iax.  In this scenario, the helium star donor is expected to survive the explosion.  We use the one-dimensional stellar evolution codes \mesa and \kepler to follow the post-impact evolution of the surviving helium companion stars.  The stellar models are based on our previous hydrodynamical simulations of ejecta-donor interaction, and we explore the observational characteristics of these surviving helium companions.  We find that the luminosities of the surviving helium companions increase significantly after the impact: They could vary from $2\mathord,500\,\Lsun$ to $16\mathord,000\,\Lsun$ for a Kelvin-Helmholtz timescale of about $10^{4}\,\yr$.  After the star reaches thermal equilibrium, it evolves as an O-type hot subdwarf (sdO) star and continues its evolution along the evolutionary track of a normal sdO star with the same mass.  Our results will help to identify the surviving helium companions of SNe Iax in future observations and to place new constraints on their progenitor models. 

\end{abstract}

\keywords{Close binary stars(254) --- Supernovae(1668) --- Type Ia supernovae(1728)}

\section{Introduction}\label{sec:intro}

Type Ia supernovae (SNe Ia) are an essential tool to measure cosmological distances and to determine cosmological parameters \citep{Riess1998AJ,Schmidt1998ApJ,Perlmutter1999ApJ}.  The progenitor systems of SNe Ia, their observational diversity and their explosion mechanisms, however, are not yet well understood \citep[see][for reviews]{Maoz2014ARAA,Maguire2017hsn,Jha2019NatAs,Soker2019NewAR}.  A plethora of theoretical progenitor models have been proposed over the past decades.  These include the single-degenerate (SD) model in which a white dwarf (WD) accretes material from a non-degenerate companion \citep{Whelan1973ApJ,Nomoto1982ApJb,Han2004MNRAS}; the double-degenerate (DD) model that involves the merger of two WDs \citep{Iben1984ApJS,Webbink1984ApJ}; the sub-Chandrasekhar mass (sub-\Mch) model in which a helium-burning or helium (He) WD companion star transfers its helium-rich material to a WD \citep{Nomoto1982ApJa,Woosley1986ApJ}; the super-\Mch model \citep{Yoon2004AAP}; the dynamically driven double-degenerate double-detonation ($\mathrm{D}^6$) model \citep{Shen2018ApJ}; the core-degenerate model \citep{Kashi2011MNRAS}; and the collisions of two WDs model \citep{Raskin2009MNRAS}.

Despite all the uncertainties in progenitor models, they still provide several ways in which they can constrain progenitor scenarios \citep[see][for a review]{Livio2018PhR}.  The most direct way to identify progenitor systems is to observe the progenitor companions in pre-explosion images because non-degenerate companions are very bright compared to a WD in the SD scenario.  There are excellent prospects to detect some of them in a pre-explosion image of nearby supernovae \citep[e.g.,][]{Li2011Nature, Kelly2014ApJ}.  During the interactions of the SN ejecta with the companion, some hydrogen-rich or helium-rich material could be stripped off the non-degenerate star.  The late-time spectra should then display the hydrogen (H) or helium features \citep[e.g.,][]{Leonard2007ApJ, Lundqvist2013MNRAS, Foley2016MNRAS, Botyanszki2018ApJ, Jacobson2019MNRAS, Magee2019AAP, Tucker2020MNRAS, Sand2021ApJ}.  The companion star survives the explosion and can be observed after the SN has faded.  Due to the mass-stripping of its outer layers and due to the shock heating during the ejecta-companion interaction, such a surviving companion star would exhibit some peculiar evolutionary features, e.g., its luminosity, effective temperature, or surface gravity, compared to those of a star with the same mass but not having experienced the impact of SN ejecta.  This makes such post-SN stars identifiable in nearby supernova remnants \citep[e.g.,][]{Ruiz-Lapuente2004Nature, Krause2008Nature, Rest2008ApJ,2012Natur.481..164S, Podsiadlowski2003,Shappee2013,Kerzendorf2014ApJ,Pan2013ApJ,Pan2014ApJ,Liu2021MNRAS,Liu2021b,Liu2022ApJ}.  Markedly, different progenitor stars would have different delay time distributions (DTDs).  This implies that comparing the DTD of progenitor models to observations can provide a decisive tool to constrain them \citep{Han2004MNRAS, Greggio2005AAP, 2009ApJ...699.2026R,Meng2010ApJ, Liu2018MNRAS, Liu2020AA}.

SNe Iax are an important sub-class of SNe Ia.  In the literature the brighter examples of this sub-class are also known as 2002cx-like events \citep{Li2003PASP, Foley2013ApJ, Jha2006AJ, Jha2017Hsn}.  Compared to `normal' SNe Ia\footnote{``Normal'' SNe Ia are defined as events for which the light curve follows the Phillips relation \citep{Phillips1993ApJ}.}, SNe Iax have a lower peak luminosity ($\mathrm{-14.2 \geqslant \mathit{M}_{\mathit{V}\!,peak} \gtrsim -18.4\,mag\,}$) and a lower ejecta velocity near the time of peak luminosity ($\mathrm{2\mathord,000 \lesssim \vert \mathit{v} \vert \lesssim 8\mathord,000\,km\,s^{-1}}$).  One of the currently widely-discussed explosion models for SNe Iax is a weak deflagration explosion of a near-Chandrasekhar mass (\Mch$\sim1.38$\,\Msun) WD \citep{Branch2004PASP,Jordan2012ApJL,Kromer2013MNRAS,Fink2014MNRAS,Lach2022AAP}.  In this explosion scenario, the \Mch\,WD is not completely disrupted by the explosion, leaving behind a WD remnant.  \citet{Kromer2013MNRAS} have shown that this weak deflagration explosion model well reproduces the observational properties of SN~2005hk, a typical SN Iax event.  \citet{Vennes2017Sci} suggest that LP~40-365 could be the WD remnant of a weak deflagration explosion due to its high space velocity and its unique atmospheric spectra \citep[see also][]{Raddi2018ApJ}.  The exact explosion mechanism of SNe Iax, however, remains under debate \citep[e.g.,][]{Stritzinger2015AAP}.

\begin{figure*}[t]
		\centering
		{\includegraphics[width=0.49\textwidth,scale=0.8]{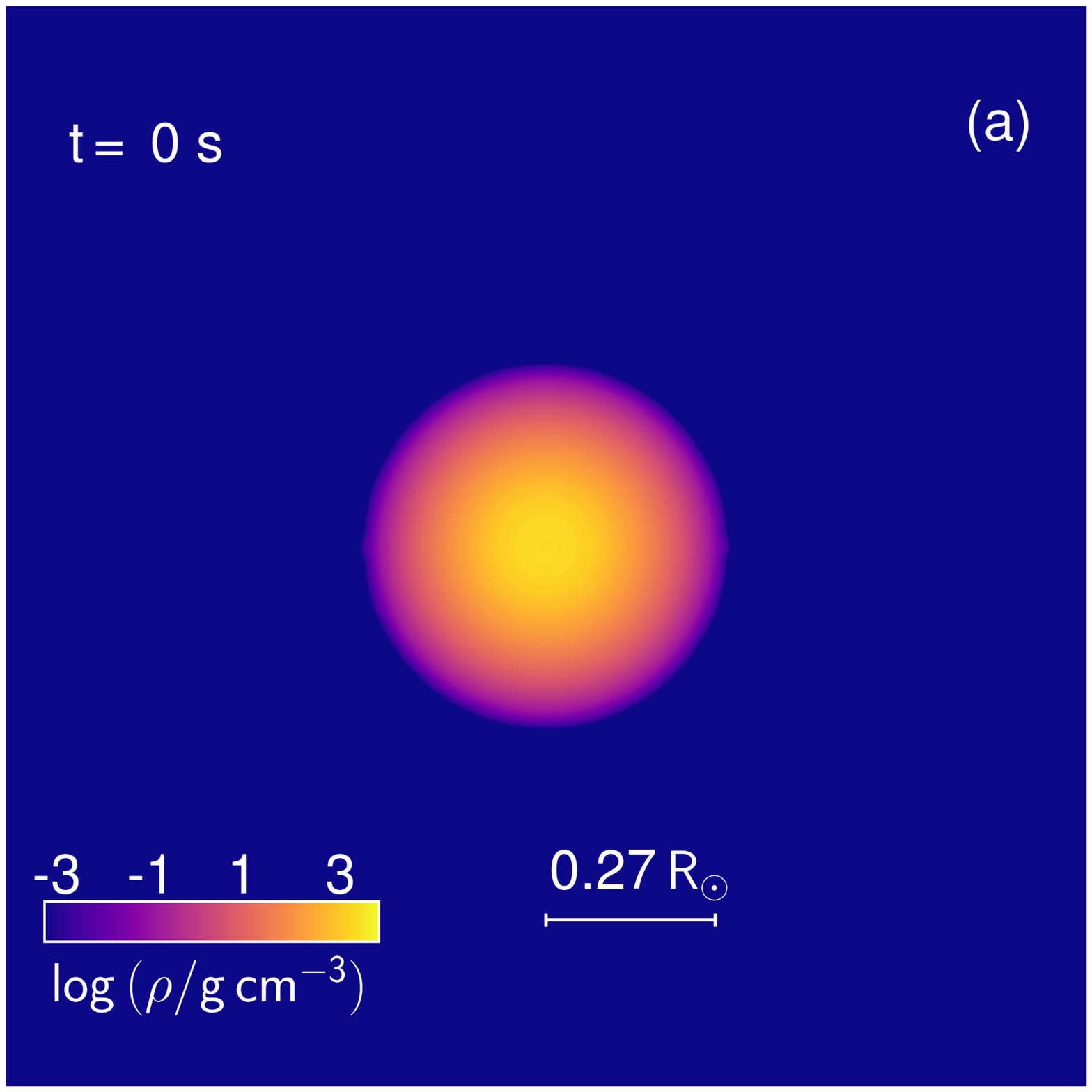}}
		\hfill
		{\includegraphics[width=0.49\textwidth,scale=0.8]{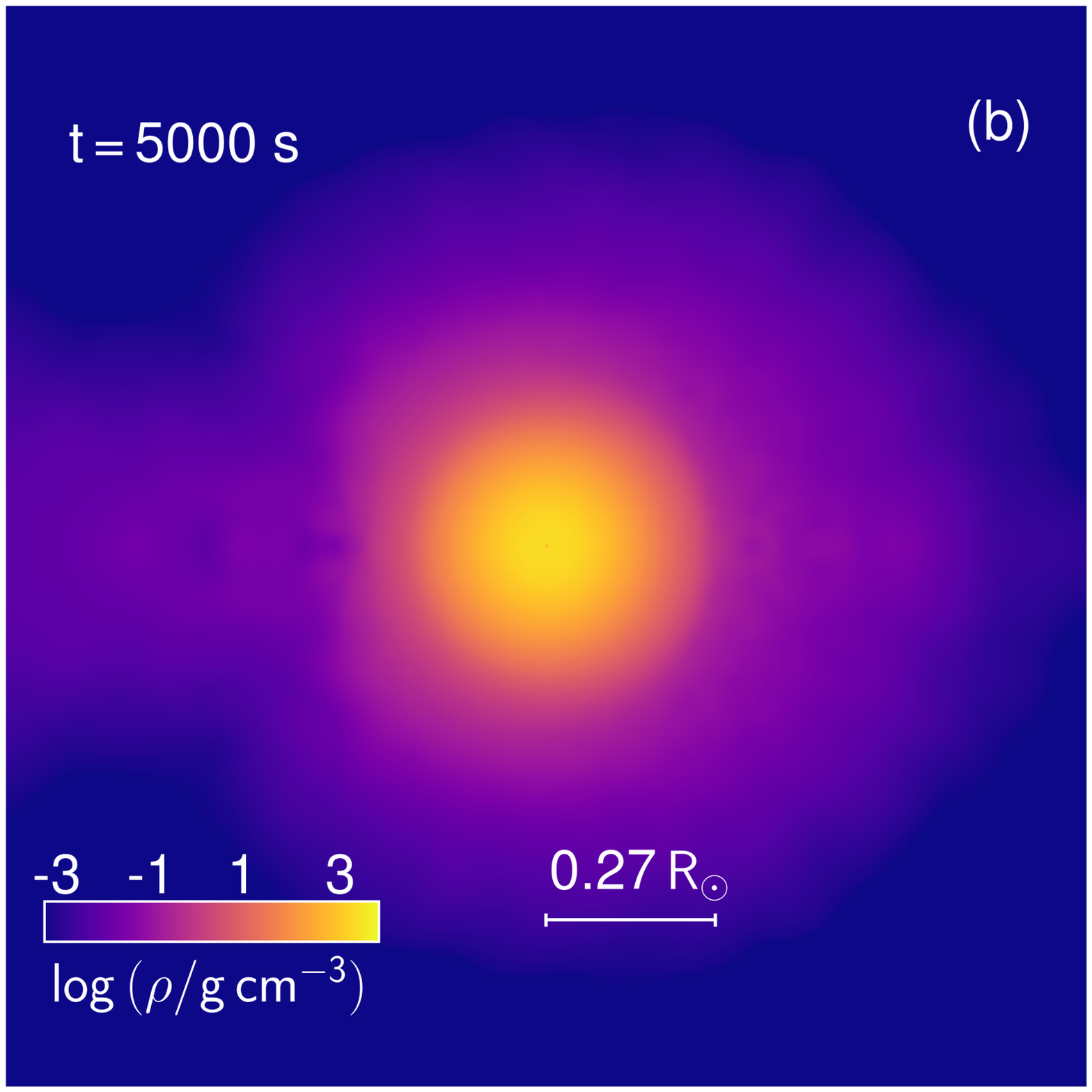}}
		\caption{\label{fig:density_slice} Density distribution of the helium-star companion at the pre-impact phase (Left-panel) and at the end (right-panel) of our SPH impact simulations for the model A0p74. In model A0p74, the He-star companion has a mass of $1.24\,\Msun$ and a radius of $0.27\,\Rsun$ at the moment of SN explosion, and the binary separation at this moment is $0.74\,\Rsun$.  Color indicates density (see color bar).}
\end{figure*}

Some observations indicate that SNe Iax may originate from progenitor systems that are composed of a WD with a He star companion.  Spectral observations revealed that the two SNe Iax, SN~2004cs and SN~2007J, have He\,\textsc{i} emission lines in their spectra \citep{Rajala2005PASP, Foley2009AJ, Foley2013ApJ, Foley2016MNRAS}.  This could be caused by a helium-star companion \citep{Foley2013ApJ}.  Most SNe Iax have been discovered in late-type galaxies with a young stellar population; the only exception is SN~2008ge that has been found in an S0 galaxy \citep{Foley2010AJ}.  This suggests that SNe Iax occur with a short delay time after star formation \citep{Lyman2013MNRAS, Lyman2018MNRAS, Takaro2020MNRAS}.  Theoretical models do predict a short DTD for SNe Type Ia for WD~+~He-star progenitor systems \citep[e.g.,][]{Liu2015AAP}, consistent with the observed DTD of SNe Iax.  Recently, \citet{McCully2014Nature} detected a bright source in pre-explosion images of the SN Iax event SN~2012Z, and this bright source has been suggested to be the helium-star companion to its progenitor system \citep{McCully2014Nature, Liu2015ApJ}.  This is the first reported pre-explosion detection of a SN Ia progenitor.

An origin of SNe Iax from WD~+~He-star progenitor systems has two general consequences:   
\begin{enumerate}
\item
The ejecta-companion interaction strips off some He material from the surface of the helium-star companion, potentially causing He lines in late-time spectra of SNe Iax.  No He lines, however, have yet been detected in the late-time spectra of SNe Iax.  This allows to estimate an upper-limit on the stripped He masses of less than about $2\times10^{-3}\,\Msun$ to $0.1\,\Msun$ \citep[e.g.,][]{Foley2013ApJ,Foley2016MNRAS,Magee2019AAP,Jacobson2019MNRAS,Tucker2020MNRAS}.  In our previous study, we have performed three-dimensional (3D) hydrodynamical simulations of the interaction between SN Ia ejecta and a helium-star companion \citep{Zeng2020ApJ} by adopting a weak deflagration explosion model of a \Mch\,WD from \citet{Kromer2013MNRAS}.  We found that only a small amount of helium, about $(4$--$7)\times10^{-3}\,\Msun$, can be stripped from the helium-star companion by SN Iax explosion.  This is consistent with the non-detection of helium emission lines in late-time spectra of SNe Iax \citep{Zeng2020ApJ}.
\item
We expect that helium-star companions in WD~+~He-star progenitor systems survive the SN Iax explosion and they are flung away with velocities that are dominated by their presupernova orbital velocities of $\sim200$--$500\,\mathrm{km\,s^{-1}}$ \citep[][]{Wang2009AA, Shen2018ApJ, Wong2019ApJ}.  A promising way to place constraints on the progenitor models of SNe Iax is to search for such surviving companion stars.  Although there are some observations that indicate the existence of a surviving companion star, to date none have been confirmed\footnote{\citet{Shen2018ApJ} identified three candidates of surviving WD companions of the $\mathrm{D}^6$ model by using astrometry from Gaia DR2.}.  For example, \citet{Foley2014ApJ} detected a red bright source about $4.1$ years after the explosion of SN~2008ha.  They argue that such a red bright source might be a surviving companion or a WD remnant as predicted by a weak deflagration explosion model.  An even more interesting example is reported by \citet{McCully2022ApJ}.  In observations of SN~2012Z taken $1\mathord,400$ days after the explosion they find a luminosity that is brighter than a normal SN Ia SN~2011fe at a similar epoch.  They suggest that a surviving companion star and a WD remnant may contribute to the extra flux of SN~2012Z in the late-time phase \citep{McCully2022ApJ}. 
\end{enumerate}

A better identification of surviving companion stars of SNe Iax after the explosion requires models for their observational properties.  In this work, we aim to provide the theoretical predictions on the post-impact characteristics of surviving helium companions for SNe Iax by mapping the results of \citet{Zeng2020ApJ} into one-dimensional (1D) stellar evolution codes.  This paper is structured as follows:  In Section~\ref{sec:method and models}, we describe our methods and initial models.  The results of post-impact evolution of surviving helium companion stars are presented in Section~\ref{sec:res}.  In Section~\ref{sec:discussion}, we compare the results with the late-time observations of SN~2012Z.  We discuss the uncertainties of our results in this section.  A summary and conclusion are given in Section~\ref{sec:summary}.

\section{Methods} \label{sec:method and models}

\begin{figure}[t]
    \centering
    \includegraphics[width=8.5cm]{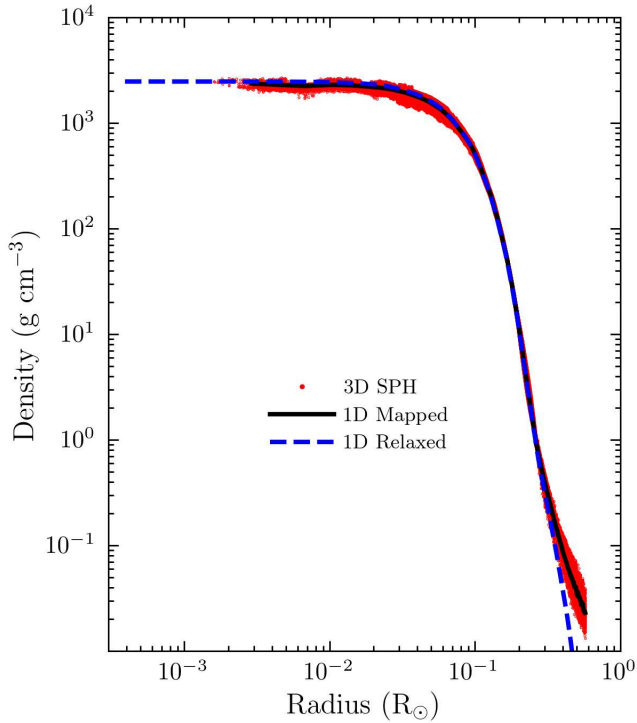}
    \caption{\label{fig:density_structure} Comparison between density distributions of the 3D SPH model (\textsl{red dots}) and radial density profile of its corresponding 1D averaged model (\textsl{black solid line}) for the A0p74 at $t=5\mathord,000\,$s after the SN explosion.  The density profile of the starting model for our \mesa calculation is shown as \textsl{blue dashed line}.  The starting model for our MESA calculation is constructed by relaxing the mapped model to hydrostatic equilibrium.}
\end{figure}

\begin{figure}[t]
    \centering
    \includegraphics[width=8.5cm]{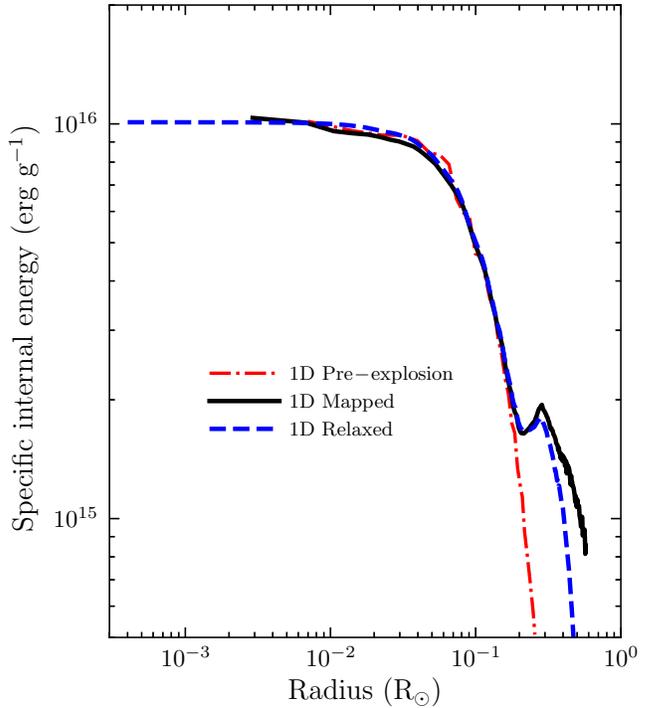}
    \caption{\label{fig:energy_structure} Comparison between radial specific internal energy profile of 1D averaged model for the A0p74 at $t=0\,$s  (\textsl{red dashed-dot line}) and at $t=5\mathord,000\,$s (\textsl{black solid line}) after the SN explosion.  The specific internal energy profile of the starting model for our \mesa calculation is shown as \textsl{blue dashed line}.  The starting model for our MESA calculation is constructed by relaxing the mapped model to hydrostatic equilibrium.}
\end{figure}

\begin{figure*}[t]
	\centering
	\includegraphics[width=1.0\textwidth]{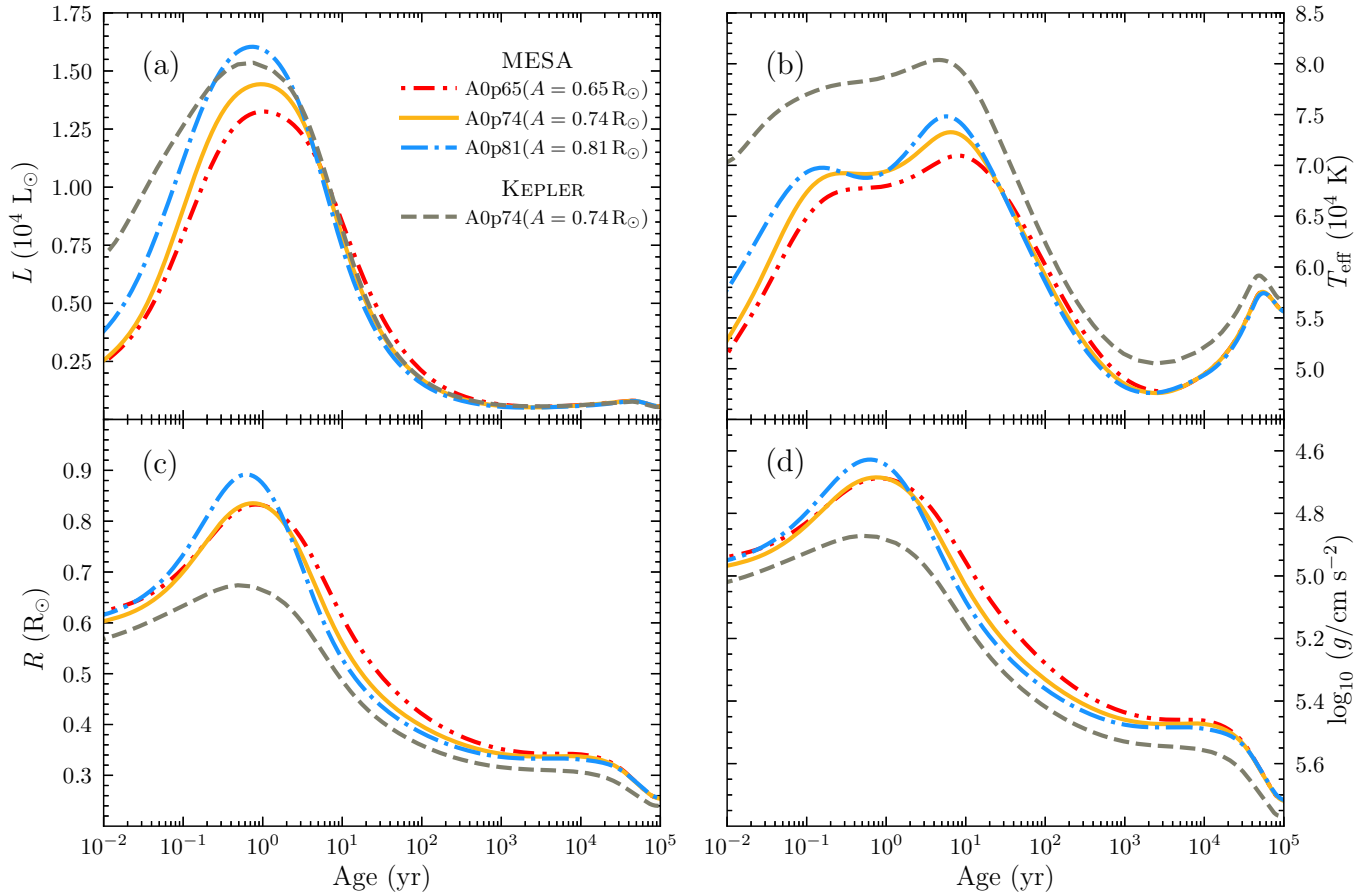}
	\caption{\label{fig:evo_ltrg} Time evolution of the post-impact luminosity, temperature, radius, and surface gravity of three surviving He companion models (A0p74, A0p65, and A0p81, see also \citealt{Zeng2020ApJ}).  For comparison the results of the post-impact evolution of Model A0p74 using the \mesa and \kepler codes are shown in \textsl{yellow solid line} and \textsl{gray dashed line}, respectively}.
\end{figure*}

\setlength{\tabcolsep}{8pt}{
	\begin{deluxetable*}{lccccclcccc}
		\centering
		\caption{Model pre- and post-explosion properties \label{tab:results}}
		%%\tablewidth{6pt}
		\tablehead{
			\colhead{$\mathrm{Name}$} & \colhead{$\mathit{A}$} & \colhead{$\mathit{M}_{2}$} & \colhead{$\mathit{R}_{2}$} & \colhead{$\log\,\mathit{L}^{\mathrm{pre}}$} & \colhead{$\log\,\mathit{T}^{\mathrm{pre}}_{\mathrm{eff}}$} & \colhead{$\mathrm{Code}$} & \colhead{$\log\,\mathit{L}^{\mathrm{peak}}$} & \colhead{$\log\,\mathit{T}^{\mathrm{peak}}_{\mathrm{eff}}$} & \colhead{$\mathit{R}^{\mathrm{peak}}$} & \colhead{$\mathit{t}^{\mathrm{peak}}$}\\
			\colhead{} & \colhead{($10^{10}\,\mathrm{cm}$)} & \colhead{$(\Msun)$} & \colhead{($10^{10}\,\mathrm{cm}$)} & \colhead{(\Lsun)} & \colhead{(\K)}& \colhead{} & \colhead{(\Lsun)} & \colhead{(\K)} & \colhead{(\Rsun)} & \colhead{(\yr)} 
		}
		%%\decimalcolnumbers
		\startdata
		A0p65 & $4.50$ & $1.24$ & $1.908$ & $2.62$ & $4.70$ & \mesa   & $4.12$ & $4.85$ & $0.83$ & $1.02$\\
		\noalign{\smallskip}
		\multirow{2}{4em}{A0p74} & \multirow{2}{1.8em}{$5.16$} & \multirow{2}{1.8em}{$1.24$} & \multirow{2}{2.4em}{$1.908$} & \multirow{2}{1.8em}{$2.62$} & \multirow{2}{1.8em}{$4.70$} & \mesa   & $4.16$ & $4.86$ & $0.84$ & $0.95$\\
		 & & & & & & \kepler & $4.19$ & $4.91$ & $0.67$ & $0.77$\\
		\noalign{\smallskip}
		A0p81 & $5.65$ & $1.24$ & $1.908$ & $2.62$ & $4.70$ & \mesa   & $4.20$ & $4.87$ & $0.89$ & $0.74$\\
		\enddata
		\tablecomments{The binary separation $(\mathit{A})$, mass $(\mathit{M}_{2})$, radius $(\mathit{R}_{2})$, luminosity $(\log\,\mathit{L}^{\mathrm{pre}})$, and effective temperature $(\log\,\mathit{T}^{\mathrm{pre}}_{\mathrm{eff}})$ of the He companion stars at the moment of SN explosion.  The $\mathit{L}^{\mathrm{peak}}$, $\log\,\mathit{T}^{\mathrm{peak}}_{\mathrm{eff}}$, and $\mathit{R}^{\mathrm{peak}}$ are the peak luminosity, the peak effective temperature, and the peak radius of post-explosion helium companion stars during the expansion phase, respectively.  The $\mathit{t}^{\mathrm{peak}}$ is the time when the maximum luminosity is reached.}
	\end{deluxetable*}
}

To investigate the observational properties of surviving helium companion stars of SNe Iax, we use the following procedure.  First, we obtain 3D surviving helium companion star models by performing hydrodynamical simulations of ejecta-companion interaction.  Second, we map the 3D structure of the surviving helium companion stars into a 1D structure.  Third, we import these mapped 1D structures into stellar evolution codes to follow their post-impact evolution.

\begin{figure*}[t]
	\centering
	\includegraphics[width=0.49\textwidth]{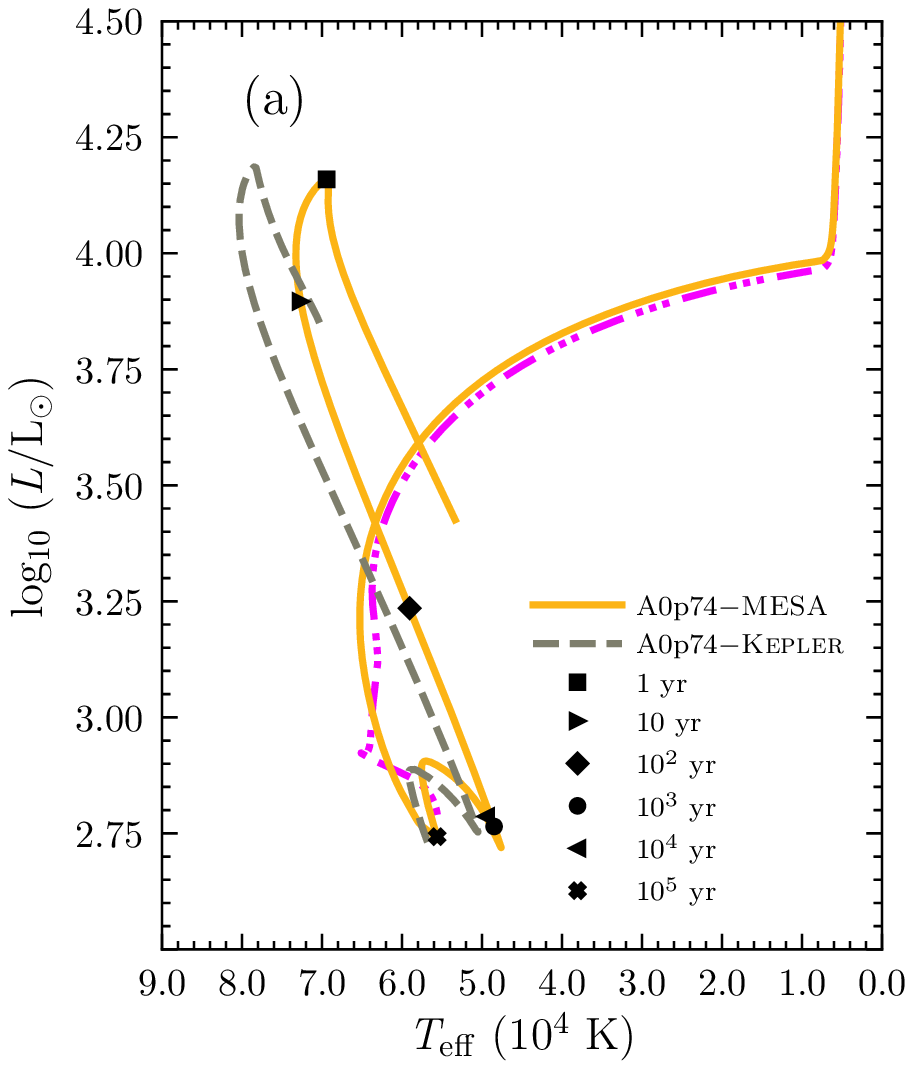}
	\hfill
	\includegraphics[width=0.49\textwidth]{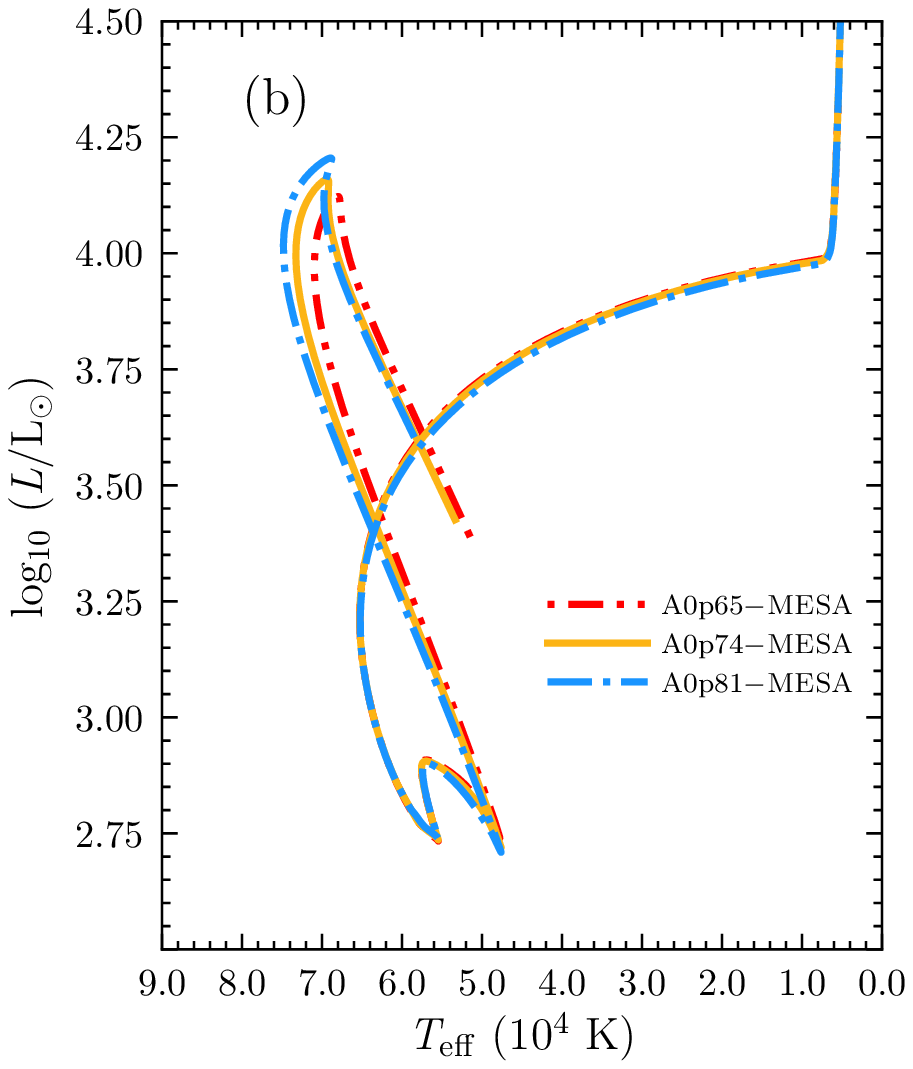}
	\caption{\label{fig:H-R} Post-impact evolution of surviving helium companion stars shown in the Hertzsprung–Russell (H-R) diagram.  \textbf{Left Panel:} a comparison of the results from \mesa (\textsl{yellow solid line}) and \kepler (\textsl{gray dashed line}) calculations for the model A0p74.  The \textsl{magenta triple-dot-dashed line} shows the evolution of a normal helium star of $1.234\,\Msun$, the same mass as A0p74 of \citet{Zeng2020ApJ}, but that does not undergo the ejecta-donor interaction.  Here, the Kepler model is only evolved for about $10^{5}\,\yr$ because it follows an evolutionary track almost identical to the MESA model after $10^{5}\,\yr$.  \textbf{Right Panel:} Evolution of models A0p65, A0p74, and A0p81 \citep[see also][]{Zeng2020ApJ} using \mesa.  For our given companion model, we set the binary separation at the moment of SN explosion to $0.65\,\Rsun$, $0.74\,\Rsun$, and $0.81\,\Rsun$, respectively}.
\end{figure*}

\subsection{3D surviving helium companion model}

In \citet{Zeng2020ApJ}, we have presented 3D smoothed particle hydrodynamics (SPH) simulations of the interaction between SN Ia ejecta and a helium companion star for SNe Iax, in which six different He-star companion models were studied. In the present work, we directly adopt three of the surviving helium companion models computed from 3D impact simulations of \citet{Zeng2020ApJ}, their models `Model~1', `Model~4', and `Model~6' (see Table~1 of \citealt{Zeng2020ApJ}) to address the post-impact evolution of such stars\footnote{For better visibility of our results (e.g., Figs.~\ref{fig:evo_ltrg} and~\ref{fig:H-R}), only three companion models of \citet{Zeng2020ApJ} are presented in this work. The post-impact observational properties of other companion models in \citet{Zeng2020ApJ} can be roughly estimated based on the results of this work.}.  Here, we rename these three selected companion models as `A0p64' (for Model~1), `A0p74' (for Model~2), and `A0p81' (for Model~3) based on their binary separations at the moment of SN explosion so that the readers can directly get the main physical differences of these models from their names. For example, `A0p74' means that the binary separation at the time of SN explosion is $A=0.74\,\Rsun$.

\emph{In the following we use `A0p74' as our reference model because the binary separation of this model in 3D simulation was adopted from the detailed binary evolution calculation.} In our detailed binary evolution calculation for model A0p74 \citep[][see their Sect.~2]{Liu2013ApJa}, an initial $1.10\,\Msun$ WD accretes helium-rich material from its $1.55\,\Msun$ companion star through Roche-lobe overflow to increase its mass to \Mch$\sim1.38$\,\Msun to trigger a thermonuclear SN explosion, which leads to a He-star companion mass, radius and binary separation at the moment of SN explosion of $M_{2}=1.24\,\Msun$, $R_{2}=0.27\,\Rsun$ and $A=0.74\,\Rsun$, respectively.  In Figure~\ref{fig:density_slice} we compare the pre- and post-impact density distribution in the orbital plane from our impact simulation for model `A0p74'. The companion star is significantly heated during the ejecta-companion interaction, leading to rapid expansion after the supernova explosion.  Due to the interaction, the surviving companion star is no longer in hydrostatic and thermal equilibrium.  A detailed description of the ejecta-companion interaction can be found in \citet{Zeng2020ApJ}.

We use two independent 1D stellar evolution codes, Modules for Experiments in Stellar Astrophysics \citep[\mesa;][]{Paxton2011ApJS, Paxton2015ApJS, Paxton2018ApJS, Paxton2019ApJS} and \kepler \citep{Weaver1978ApJ,Heger2000ApJ,Woosley2002RvMP} to follow the long-term evolution of our surviving companion stars.

\subsection{From 3D to 1D}

We map the 3D SPH surviving companion models from the impact simulations of \citet{Zeng2020ApJ} into about $200$ spherical shells.  Different properties of SPH particles within each shell, such as the density, specific internal energy, and chemical compositions, are averaged to obtain 1D radial profiles of this model \citep[see also][]{Pan2013ApJ,Liu2021MNRAS,Liu2022ApJ}.  For any property, $X$, the averaged value ($\bar{X}$) in a shell with $n$ SPH particles is given by $$\bar{X}=\frac{\sum_{i=1}^{n}\left(X_{i}\cdot m_{i}\right)}{\sum_{i=1}^{n}m_{i}}\;,$$ where $X_{i}$ is the property $X$ and $m_{i}$ is the mass of SPH particle $i$.  We use these 1D radial profiles to construct suitable starting models for the 1D post-impact evolution with the \mesa and \kepler codes.  Figure~\ref{fig:density_structure} shows a comparison between the density distribution of a 3D companion and its corresponding 1D radial profile for the A0p74.  In Figure~\ref{fig:energy_structure}, we show radial profiles of the specific internal energy of the star at its pre-explosion and post-explosion phase for the A0p74, respectively. After the explosion, energy is deposited into the companion star due to the interaction, leading to a bump of the specific internal energy in the outer layers of the post-impact star in Fig.~\ref{fig:energy_structure}.

For the long-term evolution calculations with \mesa, we use the relaxation routines provided with it \citep[][see their Appendix B for a detailed description]{Paxton2018ApJS} to construct an initial 1D surviving companion model for \mesa.  These relaxation routines are provided for constructing a model that is in hydrostatic equilibrium.  Fig.~\ref{fig:density_structure} presents the relaxed 1D density profile of the reference model computed from the relaxation routines of \mesa (blue dashed line). 

Since the starting models of our \mesa calculations are relaxed to hydrostatic equilibrium after the SN-ejecta interaction, we assess the impact of dynamical perturbations on the post-impact evolution of the surviving companion using the \kepler hydrodynamic stellar evolution code.  We directly map the 1D averaged model of SPH impact simulation into the \kepler without any relaxation process.  We then follow the long-term evolution of our reference model using \kepler for comparison.

\subsection{Input Physics}

We use the \textsc{HELM} equation of state (EOS) \citep{Timmes2000ApJS} that includes an ideal gas of ions, a Fermi–Dirac electron gas, and radiation in our \mesa calculations.  In \kepler calculations, we use the EOS from \citet{Blinnikov1996ApJS}, which is similar to the \textsc{HELM}~EOS \citep[see][for comparison of these two EOSs]{Timmes2000ApJS}.  We use the ``Type~2'' OPAL opacity table \citep{Iglesias1996ApJ} in both MESA and \kepler codes with a base metallicity of $Z_{\mathrm{base}} = 0.02$.  This opacity table respectively covers a region of the hydrogen abundance and metallicity of $0.0\leqslant X \leqslant 0.7$ and $0.0\leqslant Z \leqslant 0.1$, and it can deal with varying carbon and oxygen abundance with a specified metallicity.  We adopt a nuclear network of $21$ isotopes (\texttt{approx21.net}) for \mesa calculations.  In the \kepler code, we use the nuclear network that includes $19$ isotope (\texttt{approx-19}) ranging from the hydrogen burning to the beginning of hydrostatic silicon burning \citep{Weaver1978ApJ}.  Convection is modelled by using the mixing-length theory (MLT; \citealt{Henyey1965ApJ}) with the Ledoux criterion in both \mesa and \kepler calculations.

\section{Results from post-impact evolution models} \label{sec:res}

\subsection{Fiducial model}

\label{sec:fiducial}

Figure~\ref{fig:evo_ltrg} shows the evolution of luminosity, temperature, photosphere radius, and surface gravity of our reference model, the model A0p74, and Figure~\ref{fig:H-R} shows its evolution track in the Hertzsprung–Russell (H-R) diagram.  At the early phase of our \mesa calculations ($\sim\mathrm{10^{-2}\,\yr}$), the surviving companion star has a luminosity of $\sim2\mathord,700\,\Lsun$, an effective temperature of $\sim53\mathord,400\,\K$, a radius of $\sim0.6\,\Rsun$, and a surface gravity of $\sim\log\left(g/\mathrm{cm\,s^{-2}}\right)=4.96$.  After the SN impact, the star expands rapidly because of the energy deposited by the ejecta-companion interaction.  About one year after the explosion, the star reaches a peak radius of $\sim0.83\,\Rsun$ and a peak luminosity of $14\mathord,500\,\Lsun$ and has a surface gravity of $\sim\log\left(g/{\mathrm{cm\,s^{-2}}}\right)=4.68$.  The duration of this expanded phase is determined by the local radiative diffusion timescale \citep{Henyey1969ApJ}. As shown in Fig.~\ref{fig:evo_ltrg}b, the double bump in effective temperature is observed before the star relaxes back to thermal equilibrium. The temperature increases to about $70\mathord,000\,\K$ at $0.3\,\yr$ after the explosion and then decreases. This first temperature bump is caused by the competition between the heat transfer within the star and its expansion. At the early phase, the heat transfer is more efficient than the cooling caused by the expansion ($P\,\mathrm{d}V$ work), leading to an increase of its effective temperature as its expansion. Later, the temperature decreases because the expansion of the star starts to dominate. The effective temperature increases again as the star enters the Kelvin-Helmholtz contraction phase, leading to the second temperature bump at $\sim10\,\yr$.
The star contracts for about $10^{4}\,\yr$, and relaxes back to its thermal equilibrium state\footnote{The surviving companion star evolves on its Kelvin–Helmholtz timescale of $t_{\mathrm{KH}}=GM^{2}/RL\sim10^{4}\,\yr$ \citep[][]{Marietta2000ApJS}.  Here, $R$, $M$, and $L$ are the radius, mass, and luminosity of the surviving companion star; $G$ is the gravitational constant.}, becoming an O-type hot subdwarf (sdO)-like object \citep[see also][]{Pan2013ApJ}.  At about $10^{5}\,\yr$ after the impact, a second local `peak' of luminosity is observed in the Fig.~\ref{fig:H-R} because of the onset of central helium-burning.  The star then evolves in the H-R diagram similar to a regular sdO-like star (triple-dot-dashed line in Fig.~\ref{fig:H-R}).  Our post-impact evolution calculations are terminated at $\sim10^{6}\,\mathrm{yr}$ because after that the surviving companion models follow almost the same evolutionary track, that of a regular sdO-like star.

\begin{figure*}[t]
	\centering
	\includegraphics[width=0.49\textwidth]{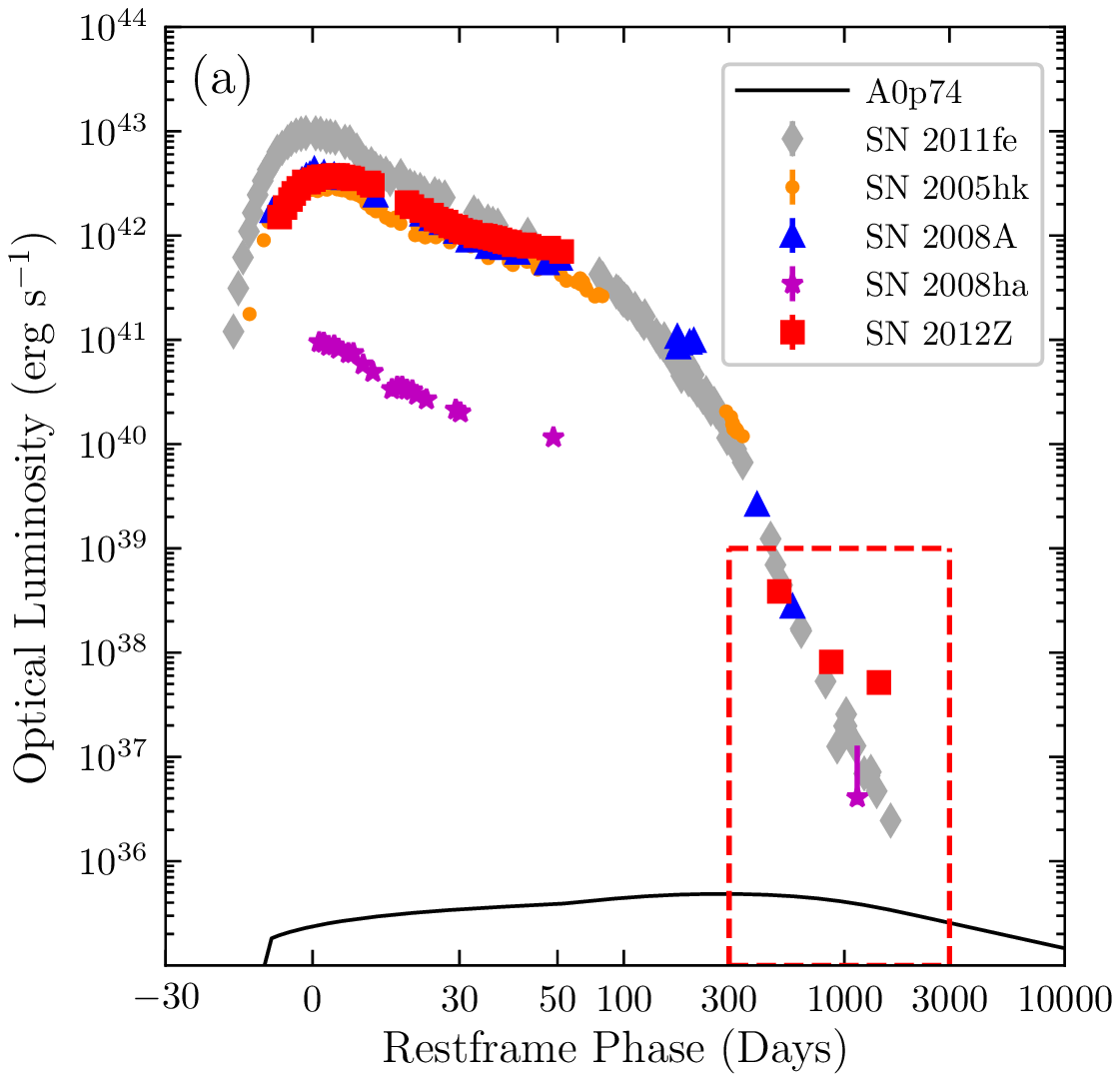}
	\hfill
	\includegraphics[width=0.49\textwidth]{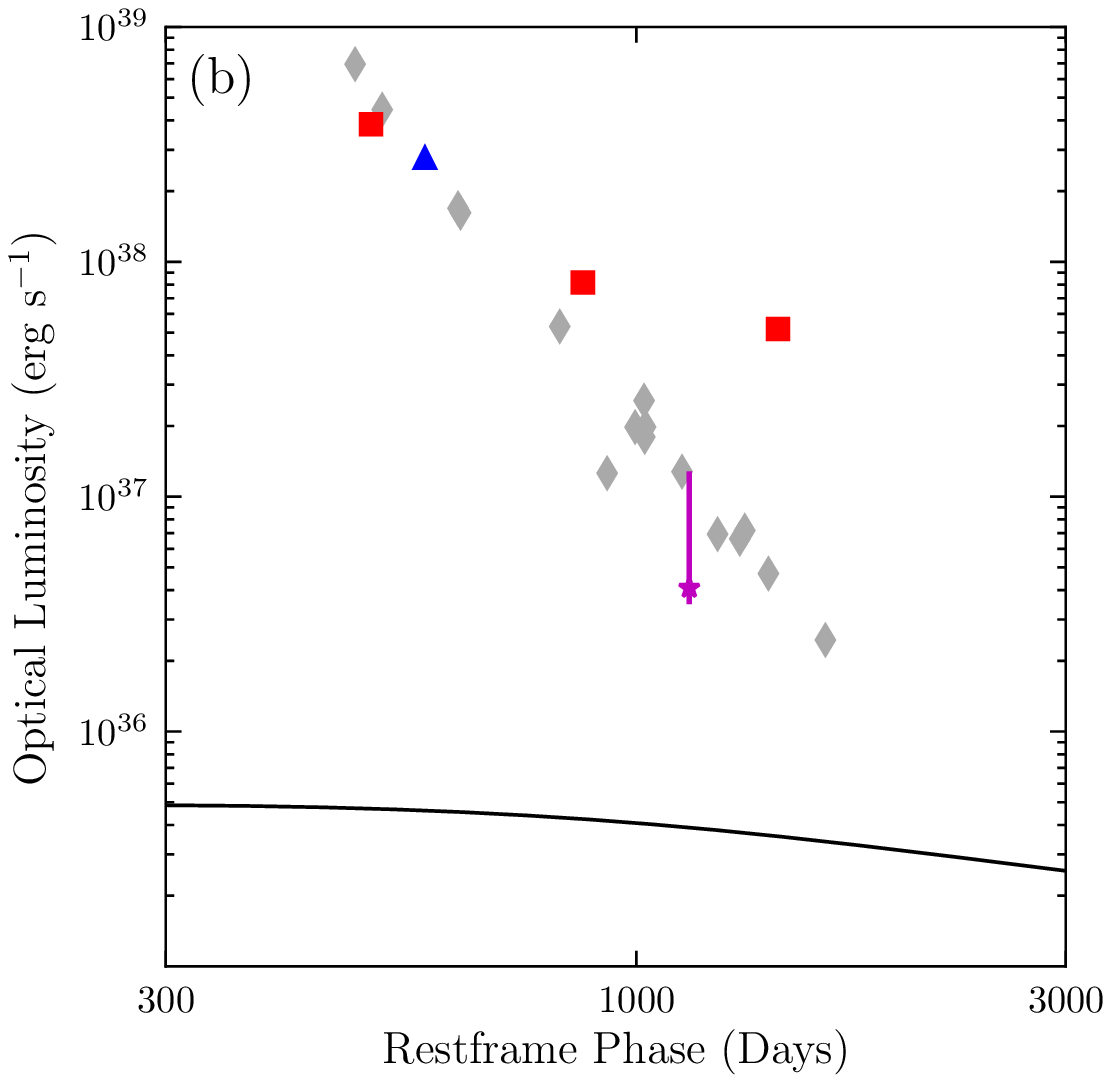}
	\caption{\label{fig:LC} A comparison between model A0p74 (\textsl{black solid line}) and the observed optical light curves of SNe~2005hk (\textsl{orange point}), 2008A (\textsl{blue up triangle}), 2008ha (\textsl{magenta star}), 2011fe (\textsl{grey thin diamond}), and 2012Z (\textsl{red square}).  \textbf{Left Panel:}  Light curves of these supernovae in the early phase of the explosion and in the corresponding late phase.  \textbf{Right Panel:} Enlarged image of the data in the \textsl{red dashed box}.}
\end{figure*}

To test the role of hydrodynamical imbalance of the post-impact companion star and its evolution we follow the long-term evolution of A0p74 by using the \kepler code, as described in Section~\ref{sec:method and models}.  The results are shown as grey dashed lines in Figs.~\ref{fig:evo_ltrg} and~\ref{fig:H-R}.  The post-impact evolution tracks in the H-R diagram of both models are similar. Especially, about $10^{4}\,\yr$ after the explosion, the \kepler model follows an evolution track almost identical to the \mesa model and becomes an sdO-like star.  This suggests that the approximate hydrostatic initial setup generated by \mesa should be a close approximation for the post-impact evolution on this longer time scale.  However, some differences in the post-impact properties between \kepler model and \mesa model are still observed before the star has relaxed back into thermal equilibrium, although the differences will not alter the main conclusions of this work.  For instance, the \kepler model reaches a higher luminosity and effective temperature, and a smaller radius than the \mesa model.  To avoid both artifacts from rezoning and potential inconsistencies from trying to construct a hydrostatic stellar atmosphere on top of the mapped hydrodynamic model, the \kepler model kept the initial resolution of the mapped grid.  Such a \emph{much} cruder surface resolution of the \kepler model could raise some differences in the results.

\subsection{Dependency on the presupernova binary separation}

To investigate the impact of the pre-explosion binary separation on the ejecta-companion interaction, \citet{Zeng2020ApJ} performed a set of impact simulations (see their Table~1).  For a given companion model, they artificially changed the ratio of the binary separation to the companion radius at the time of the SN Ia explosion within the range from $2.36$ to $2.96$.  All other parameters were kept the same \citep{Liu2013ApJa}.

We selected two models of \citet{Zeng2020ApJ}, `A0p65' and `A0p81', to explore the influence of presupernova binary separation on the post-impact evolution of a surviving companion star.  Our models A0p65 and A0p81 have a value of $A/R_{2}$ of $2.36$ and $2.96$, respectively.  Figs.~\ref{fig:evo_ltrg} and~\ref{fig:H-R} show their post-impact evolution.  Compared to the reference model ($A/R_{2}=2.70$), we find that a larger presupernova binary separation results in a higher peak luminosity during the expansion phase.  This may be strongly dependent on a balance between the amount of deposited energy and the depth of energy deposition caused by the shock heating during the ejecta-companion interaction.  For a given companion star, a larger binary separation reduces the cross-sectional area of the ejecta-companion interaction and thus reduces the amount of energy deposition.  The depth of the energy deposition, however, is shallower in this case, allowing it to be emitted quicker.  This results in a higher peak luminosity of the post-supernova companion star despite the lower energy deposit.  Table~\ref{tab:results} lists model properties and key results.

\section{Discussion} \label{sec:discussion}

\begin{figure*}[t]
	\centering
	\includegraphics[width=1.0\textwidth]{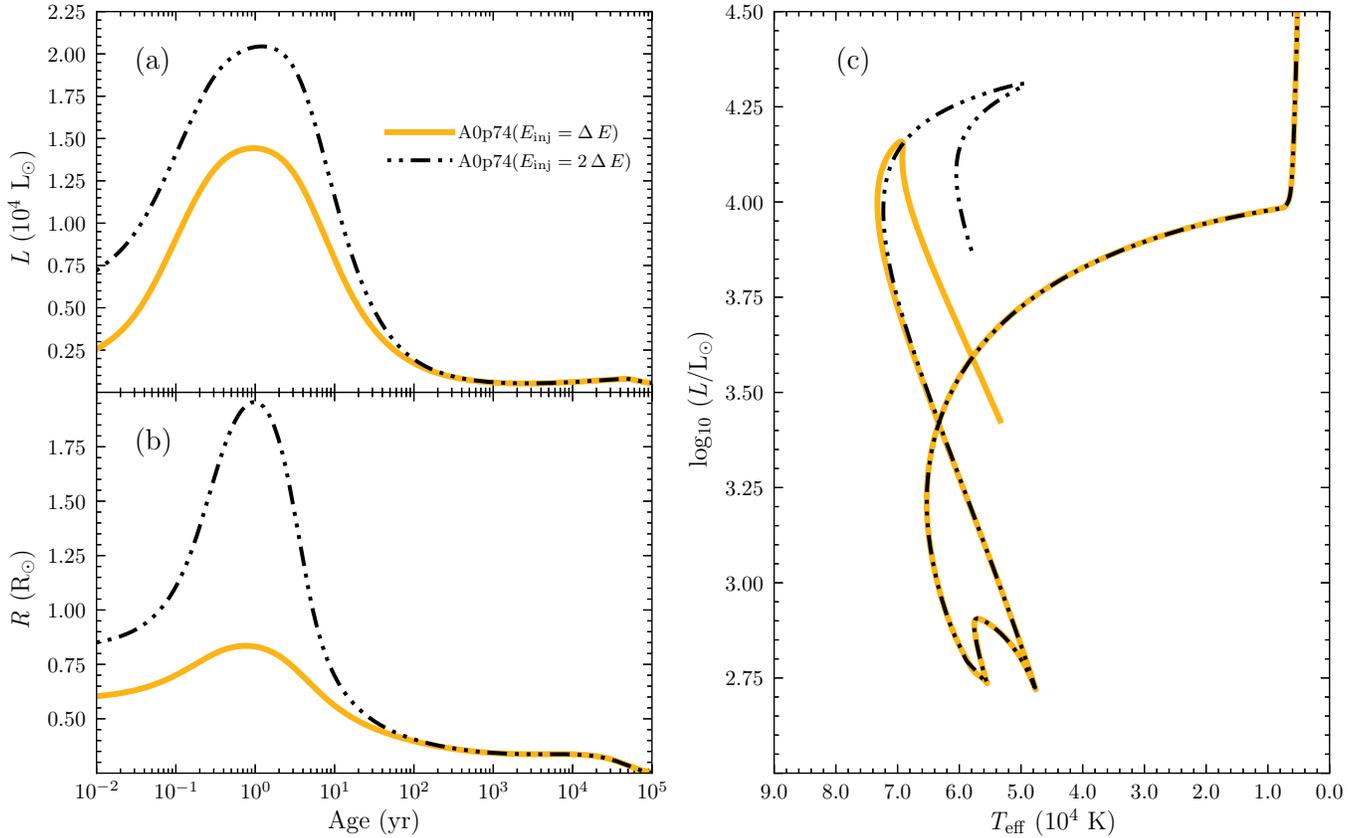}
	\caption{\label{fig:evo_ltrg_test}  Similar as Figs.~\ref{fig:evo_ltrg} and~\ref{fig:H-R}, but for the test by artificially increasing the deposited energy ($\Delta E$) by a factor of two for a given companion model A0p74.}
\end{figure*}

\subsection{Comparing with the observations of SN~2012Z}

\citet{McCully2022ApJ} recently published extremely late-time photometry data of SN~2012Z, taken with Hubble Space Telescope (\textsc{HST}) at about $1\mathord,400$ days past maximum light.  They find that at this epoch the luminosity of SN~2012Z is brighter than a normal SN Ia, SN~2011fe, by a factor of two, although at peak SN~2011fe is a factor of two brighter than SN~2012Z.  Based on standard SN Ia light curves, this suggests that there is an extra flux contribution besides SN itself.  \citet{McCully2022ApJ} have further suggested that either a surviving companion star or a remnant of the exploding WD in a weak deflagration model \citep{Shen2007ApJ}, or the combination of both may be the contributor.

In Figure~\ref{fig:LC}, we compare our results to this late-time observation of SN~2012Z by \citet{McCully2022ApJ} to test whether our surviving helium companion models can provide an explanation for the origin of extra flux observed in SN~2012Z.  To obtain the optical luminosity of our surviving companion model, we integrate the flux with a wavelength range of $340$--$970\,\mathrm{nm}$ by assuming a black body model with an effective temperature and photosphere radius given by our surviving helium companion model.  Figure~\ref{fig:LC} shows that our surviving companion star cannot be the main contributor of the late-time extra flux observed in SN~2012Z.  Its luminosity is much too low.  At $1\mathord,400$ days after the SN, our surviving companion star has a high surface temperature of $47\mathord,000\,\mathrm{K}$ and emits most of its flux in the UV rather than in the optical frequency band.

\subsection{Model uncertainties}

In their 3D impact simulations, \citet{Zeng2020ApJ} only used the so-called `N5def model' of \citet[][]{Kromer2013MNRAS} as their SN Iax explosion model.  It is a pure weak deflagration explosion model of a Chandrasekhar-mass WD that reproduces the observational features of SN~2005hk \citep{Kromer2013MNRAS}.  A wide range of observed peak luminosities of SNe Iax, however, indicates that SNe Iax might span a wide range of explosion energies.  For example, \citet{Kromer2015MNRAS} show that an explosion model with a low explosion of $1.8\times10^{48}\,\mathrm{erg}$ -- about two orders of magnitude lower than that of the N5def model -- is needed to explain the faint SN Iax event SN~2008ha.  A higher or lower explosion energy could affect the ejecta-companion interaction and thus the post-impact evolution and properties of the surviving companion star \citep{Liu2013ApJa, Liu2022ApJ,Pan2013ApJ}.  The exact explosion mechanism that leads to SNe Iax, however, is still unknown. \citet{Lach2022AAP} find that the faint part of the SN~Iax class is difficult to reconcile with the Chandrasekhar-mass deflagration model.  A pulsational delayed detonation model has been proposed for SNe Iax \citep{Hoeflich1995ApJ, Stritzinger2015AAP}.  These uncertainties in the nature of SNe Iax and their explosion models translates into uncertainties of the model presented here in reproducing specific supernova observations that may have different supernova energies. To simply test the influence of explosion energy on the post-impact evolution of the star, we artificially increase the amount of energy deposition by a factor of two based on the model A0p74 for a given depth of energy deposition.  As a consequence, we find that the post-impact maximum luminosity and radius of the star respectively increase by a factor of $\sim1.5$ and $\sim2.4$ compared with those of A0p74 model (see Figure~\ref{fig:evo_ltrg_test}).  In addition, the peak effective temperature decreases to $49\mathord,000\,\K$ (which is $20\mathord,000\,\K$ lower than that of A0p74 model), suggesting that it will emit more flux in optical frequency band than A0p74 model. In a realistic case, however, a higher explosion energy would not only lead to a higher amount of energy deposition but also to a strong mass-stripping and a deeper energy deposition.  To reach a definite conclusion, a future study is still required to perform a 3D impact simulation with different explosion models.

In this work, we assume that SNe Iax are due to a weak deflagration explosions in a binary progenitor systems composed of a WD and a helium star donor.  Our computational resources only allowed us a limited study that includes models with different surviving helium companion star separations and only for one helium companion star mass.  The nature of the SNe Iax progenitor systems, however, remains an open question.  Even for the same explosion mechanism that we assume here, the companion star could have been a main sequence (MS) star.  \citet{Liu2021MNRAS} have investigated the long-time evolution of surviving MS companion stars.  They find that due to shocking heating during the interaction the surviving MS companion stars could have luminosities in the range $5$--$1\mathord,000\,\Lsun$ for $\sim1\mathord,000$ years after the supernova.  In the future, a more comprehensive study of progenitor systems and supernova explosion models is to better cover the full range of potential properties of surviving companion stars of SNe Iax for comparison with observational data.

Some radioactive heavy isotopes such as $^{56}\mathrm{Ni}$ from SN ejecta could be captured by the companion star during the ejecta-companion interaction \citep[e.g.,][]{Liu2012AAP,Liu2013ApJa,Pan2012ApJ}.  The effect due to heating by these captured radioactivities on the post-impact evolution of the surviving companion star is not considered in this work.  The decay of captured radioactive isotopes is expected to heat the companion star during its post-impact evolution \citep{Shen2017ApJ}. \citet{Pan2013ApJ} have investigated the effect of the nickel contamination on the post-impact properties of a surviving helium-star companion for normal SNe Ia.  They find that the inclusion of nickel contamination does not affect the post-impact luminosity of the star significantly.  However, this leads to a potential increase of the star by a factor of up to ten because of the increase of opacity in the heated outer layers of the star due to the deposited nickel.  The contamination of radioactive heavy elements could alter the observational properties of our surviving companion star and will be explored in future work.

\section{Summary and Conclusion} \label{sec:summary}

We map three surviving helium companion models of \citet{Zeng2020ApJ} into the 1D stellar evolution codes \mesa \citep{Paxton2011ApJS, Paxton2018ApJS} and \kepler \citep{Weaver1978ApJ,Heger2000ApJ,Woosley2002RvMP}.   These initial models originate from 3D hydrodynamical simulations of the ejecta-companion interaction for SNe Iax.   We follow their post-impact evolution for more than $10^6\,\yr$ after the explosion.  Our results and conclusions are summarized as follows:

\begin{itemize}

\item We find that an energy deposition into the surviving companion star due to the shock heating during the ejecta-companion interaction causes a rapid increase of the luminosity of the star after the supernova impact.  About one year after the explosion, the star reaches a peak luminosity of $\sim10^{4}\,\Lsun$, radiating away the deposited energy.

\item Subsequently, the surviving helium companion stars contract and return to a thermal equilibrium state at about $10^{4}\,\yr$ after the supernova impact.  They then evolve into sdO-like stars due to the release of gravitational energy \citep[see also][]{Pan2013ApJ,Liu2021b,Liu2022ApJ}.  As a result, the surviving helium companion stars of SNe Iax might contribute to the formation of single sdO stars.

\item For a given helium companion star and explosion model, we find that a wider pre-explosion binary separation leads to a higher peak luminosity before the star starts to reestablish thermal equilibrium.

\item Comparing our results to the late-time \textsc{HST} observations of SN~2012Z taken about $1\mathord,400$ days after the explosion \citep{McCully2022ApJ}, we find that with the specific system and supernova parameters used in this work for the surviving companion models, we fail to explain the late-time observation of SN~2012Z in the optical.  Our models are fainter and emit most of their flux in the UV at that epoch.  

\item A larger library of models with different system parameters and supernova models and energies is needed in the future to allow better identification of surviving companion stars of SNe Iax by future observations, in particular, as we should expect to find a plethora of SN Iax candidates with the upcoming Vera Rubin Observatory.

\end{itemize}

\begin{acknowledgments}

We thank the anonymous referee for his/her helpful comments.  We thank Xiangcun Meng, Hailiang Chen, and Jiajia Li for their useful discussions.  Y.T.Z.\ would like to thank Yangyang Zhang and Heran Xiong for their helping of using the \mesa.  This work is supported by the National Natural Science Foundation of China (NSFC, Nos.~11873016, and 11733008), the Chinese Academy of Sciences, the National Key R\&D Program of China (Nos.~2021YFA1600400 and 2021YFA1600401), and Yunnan Province (Nos.~12090040, 12090043, 202001AW070007, 2019HA012, and 2017HC018).  A.H.\ was supported, in part, by the Australian Research Council (ARC) Centre of Excellence (CoE) for Gravitational Wave Discovery (OzGrav) through project number CE170100004, by the ARC CoE for All Sky Astrophysics in 3 Dimensions (ASTRO 3D) through project number CE170100013, and by the National Science Foundation under Grant No.~PHY-1430152 (JINA Center for the Evolution of the Elements, JINA-CEE).  The work of F.K.R.\ is supported by the Klaus Tschira Foundation and by the Deutsche Forschungsgemeinschaft (DFG, German Research Foundation) -- Project-ID 138713538 -- SFB 881 (``The Milky Way System'', Subproject A10).  The authors gratefully acknowledge the ``PHOENIX Supercomputing Platform'' jointly operated by the Binary Population Synthesis Group and the Stellar Astrophysics Group at Yunnan Observatories, Chinese Academy of Sciences.  This work made use of the Heidelberg Supernova Model Archive \citep[HESMA,][see \url{https://hesma.h-its.org}]{Kromer2017MmSAI} 

\end{acknowledgments}

\software{\mesa \citep[version 12778;][]{Paxton2011ApJS, Paxton2015ApJS, Paxton2018ApJS, Paxton2019ApJS}, \kepler \citep{Weaver1978ApJ,Heger2000ApJ,Woosley2002RvMP}, yt-project \citep[version 3.4.1;][]{Turk2011ApJS}, matplotlib \citep[version 3.1.0;][]{Hunter2007CSE}, numpy \citep[version 1.16.4;][]{Van2011CSE} and scipy \citep[version 1.1.0;][]{Virtanen2020NatMe}}

%\bibliography{ref}{}
%\bibliographystyle{aasjournal}

\end{document}